\documentclass[preprint,12pt,authoryear]{elsarticle}
\usepackage[a4paper,margin=1in]{geometry}   % standard narrow margins

\usepackage{microtype}              % enables font expansion & protrusion
\usepackage[Export]{adjustbox}

\usepackage{listings}
\usepackage{caption}
\usepackage{booktabs,tabularx}

\usepackage{amsmath,graphicx}
\usepackage[hidelinks]{hyperref}  % makes \ref clickable in the PDF
\usepackage{lineno}

\usepackage{algorithm}
\usepackage{algpseudocode}
\modulolinenumbers[5]

\usepackage{etoolbox}

\begin{document}
\begin{frontmatter}
  \title{NeuroGaze: A Hybrid EEG and Eye-Tracking Brain-Computer Interface for Hands-Free Interaction in Virtual Reality}
  \author[ucf]{Kyle Coutray}
  \author[ucf]{Wanyea Barbel}
  \author[ucf]{Zack Groth}
  \author[ucf]{Joseph J. LaViola Jr\corref{cor1}}

  \address[ucf]{Interactive Systems and User Experience Research Cluster, Department of Computer Science, University of Central Florida, Orlando, FL, USA}

  \cortext[cor1]{Corresponding author}

  \begin{abstract}
  Brain-Computer Interfaces (BCIs) have traditionally been studied in clinical and laboratory contexts, but the rise of consumer-grade devices now allows exploration of their use in daily activities. Virtual reality (VR) provides a particularly relevant domain, where existing input methods often force trade-offs between speed, accuracy, and physical effort. This study introduces NeuroGaze, a hybrid interface combining electroencephalography (EEG) with eye tracking to enable hands-free interaction in immersive VR. Twenty participants completed a 360° cube-selection task using three different input methods: VR controllers, gaze combined with a pinch gesture, and NeuroGaze. Performance was measured by task completion time and error rate, while workload was evaluated using the NASA Task Load Index (NASA-TLX). NeuroGaze successfully supported target selection with off-the-shelf hardware, producing fewer errors than the alternative methods but requiring longer completion times, reflecting a classic speed-accuracy tradeoff. Workload analysis indicated reduced physical demand for NeuroGaze compared to controllers, though overall ratings and user preferences were mixed. These findings demonstrate the feasibility of hybrid EEG+gaze systems for everyday VR use, highlighting their ergonomic benefits and inclusivity potential. Although not yet competitive in speed, NeuroGaze points toward a practical role for consumer-grade BCIs in accessibility and long-duration applications, and underscores the need for improved EEG signal processing and adaptive multimodal integration to enhance future performance.

  \end{abstract}
  \begin{keyword}
    Brain-Computer Interface (BCI) \sep Electroencephalography (EEG) \sep Eye tracking \sep Virtual reality (VR) \sep Hybrid interfaces \sep Hands-free interaction \sep Human-computer interaction (HCI) \sep Accessibility
  \end{keyword}
\end{frontmatter}
\section{Introduction}
Virtual reality (VR) systems have advanced rapidly in terms of visual immersion and motion tracking, but interaction remains a central challenge \citep{1011452,laviola2017book,101063}. The input devices that mediate user actions fundamentally shape the quality of the experience, and each current method carries trade-offs. Handheld controllers remain the most common solution, offering speed and precision through ray casting and button presses. However, extended use of controllers can be fatiguing, particularly in tasks that require repetitive pointing or in scenarios where hands-free interaction is desirable \citep{9419261}. Gaze-based dwell selection has been proposed as a more natural, ergonomic alternative \citep{9873986} where users fixate on a target and selection occurs after a brief dwell time, building on decades of research into the fundamental dynamics of eye movements \citep{101167,101001}. While hands-free, this approach is slower, vulnerable to the “Midas touch” problem \citep{Tang02062025} of unintended activations, and can feel unnatural when prolonged fixations are required \citep{8699248,isomoto,Chakraborty}. More recently, combinations of eye tracking with manual gestures, such as pinch confirmation, have improved speed and reduced false selections \citep{Zhang02072020,Vertegaal,Stellmach}. Yet these methods still depend on reliable hand mobility and introduce motor demands that limit accessibility for some users \citep{Gherman}. 

Brain-Computer Interfaces (BCIs) offer an intriguing pathway to augment VR interaction by providing a neural channel for intent confirmation \citep{Saxena}. Traditionally restricted to clinical and tightly controlled experimental contexts \citep{s19061423,s120201211}, BCIs are now becoming accessible outside the lab with the rise of consumer-grade headsets and biosensors. These devices make it feasible to test interaction techniques not only in laboratory studies but also in the context of daily activities \citep{Vasiljevic20012020,Pan2017EvaluationOC,a10795393}. Prior research has demonstrated the feasibility of integrating electroencephalography (EEG) with eye tracking for selection tasks in desktop environments and experimental prototypes \citep{Putze,Putze2,Hild}. For example, hybrid gaze+EEG systems have been used to disambiguate visual targets, detect covert attention, or reduce false activations \citep{Shishkin,kalaganis2018erroraware,Vortmann,evain}. However, most of this work has remained confined to controlled laboratory setups or 2D displays, with limited exploration in fully immersive VR environments \citep{larsen2024synchronized} and little emphasis on consumer-grade hardware. As a result, the real-world practicality of such systems for daily activities remains uncertain.

To address this gap, this study introduces and evaluates NeuroGaze, a hybrid EEG and eye-tracking interface designed for immersive VR using readily available consumer devices (Meta Quest Pro for eye tracking and Emotiv EPOC X for EEG). Unlike prior work that has focused narrowly on proof-of-concept demonstrations, we directly benchmark NeuroGaze against two widely adopted VR input methods: hand controllers and eye tracking with pinch gestures. In doing so, we provide the first comparative validation of a consumer-grade hybrid EEG+gaze system in immersive VR. Our evaluation maps the trade-offs between speed, accuracy, and physical effort across these modalities, situating NeuroGaze within the broader design space of VR interaction. The findings reveal both the potential and the current limitations of hybrid BCIs for daily activities, highlighting their promise as an accessible, ergonomic alternative for users who may benefit from hands-free, low-effort interaction.

\section{Materials and Methods}

\subsection{Participants}
Twenty healthy adult volunteers (12 male, 8 female; age range 18-32 years) were recruited from the university community. All participants reported normal or corrected-to-normal vision, no history of neurological or motor impairments, and no susceptibility to simulator sickness. Inclusion criteria required participants to be at least 18 years of age, proficient in English, and physically able to wear both the EEG headset and the VR head-mounted display.

Participants represented a broad range of prior VR experience, from no exposure to frequent recreational use. Approximately 25\% of the sample reported little or no prior VR experience, 60\% reported moderate to above-moderate experience, and 15\% described themselves as very experienced. Comparable distributions were observed for AR exposure and VR gaming, indicating that the sample encompassed both novices and highly experienced users. %(Table \ref{tab:responses}).

All participants provided written informed consent prior to participation. The study was approved by the university's Institutional Review Board (IRB ID: STUDY00006401).

\subsection{Apparatus}
\begin{figure}[ht]
\begin{center}
\includegraphics[width=15cm]{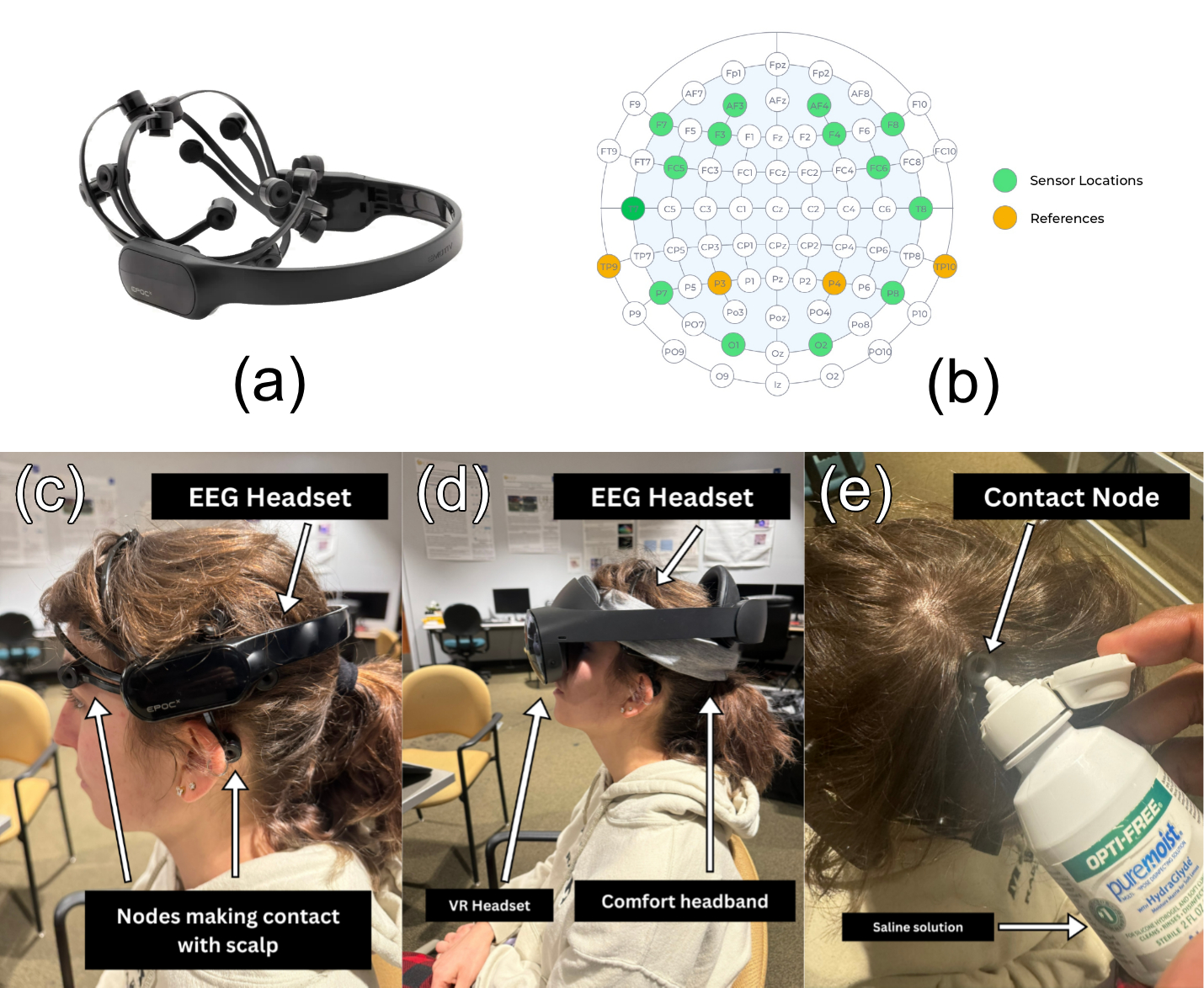}
\end{center}
\caption{(a) The Emotiv EPOC X headset used for EEG data collection. (b) Electrode montage showing the 14 active sensor locations (green) and mastoid reference electrodes (orange) based on the international 10-20 system. (c) Emotiv EPOC X EEG headset with electrodes in contact with the scalp. (d) Combined configuration of the EPOC X, comfort headband, and Meta Quest Pro VR headset worn simultaneously. (e) Application of saline solution to EPOC X electrodes to maintain stable contact quality.}\label{fig:epocsetup}
\end{figure}

The immersive environment was presented using a Meta Quest Pro head-mounted display (Meta Platforms Inc., USA) with integrated binocular eye tracking. The headset provided real-time gaze vectors at a sampling rate of 72 Hz \citep{hou}, and participants completed a standard five-point calibration at the beginning of each session. EEG activity was recorded using an Emotiv EPOC X headset (Emotiv Inc., USA), which features 14 active electrodes positioned according to the international 10-20 system \citep{khazi2012eeg} (AF3, F7, F3, FC5, T7, P7, O1, O2, P8, T8, FC6, F4, F8, AF4) with mastoid references (TP9, P3, P4, TP10) shown in Figure \ref{fig:epocsetup}A-B. EEG signals were captured internally at 2048 Hz, then downsampled to 128 Hz for wireless transmission via Bluetooth Low Energy \citep{emotiv2020epocx}. Electrode-skin contact quality was continuously monitored, with saline solution (OPTI-FREE) reapplied as needed to maintain stable impedance. The NeuroGaze setup required participants to wear both the Emotiv EPOC X and Meta Quest Pro simultaneously, often secured with a comfort headband to ensure reliable electrode contact (Figure \ref{fig:epocsetup}C-E).

\begin{figure}[ht]
\begin{center}
\includegraphics[width=17cm]{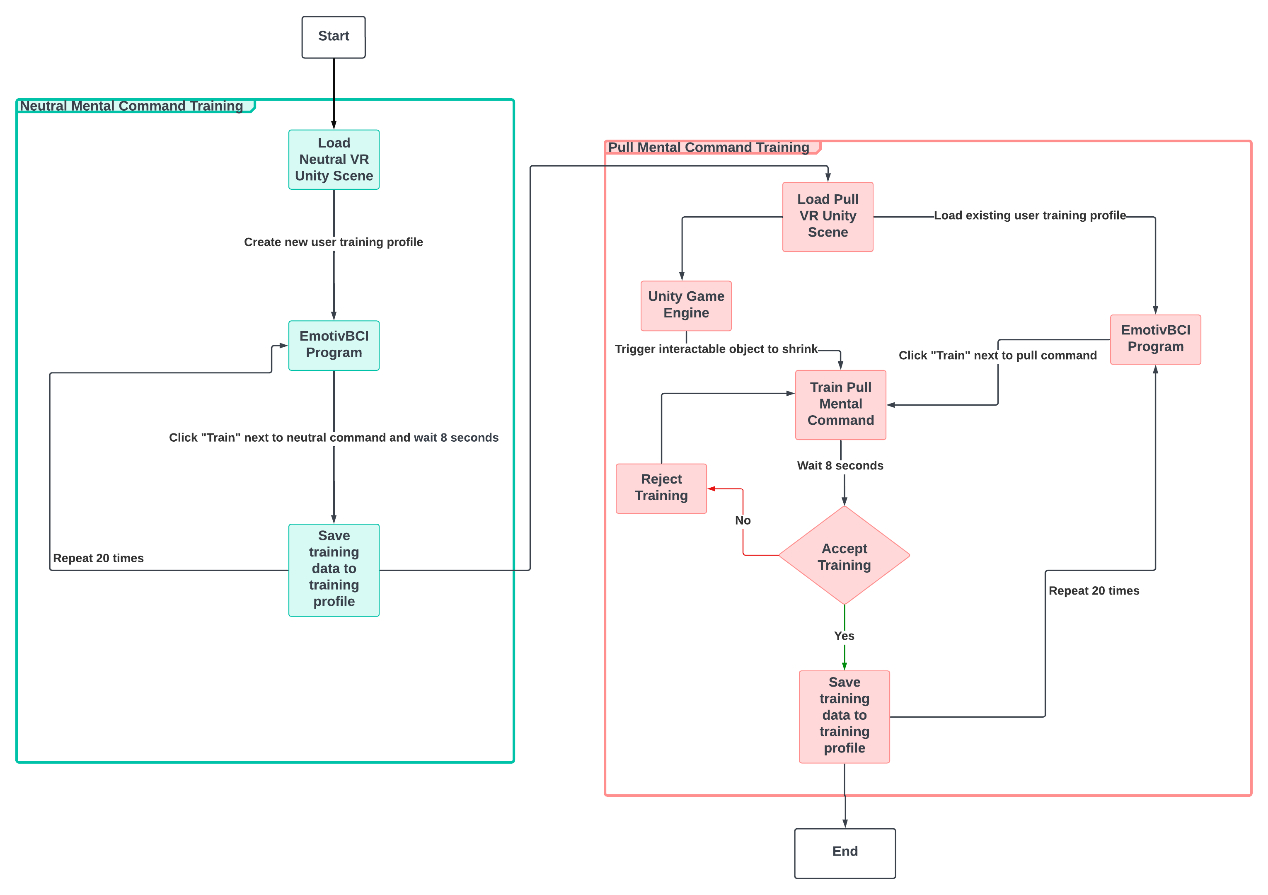}
\end{center}
\caption{Flowchart of the NeuroGaze EEG calibration procedure. The process consists of two stages: (left, teal) neutral mental command training, where participants repeatedly train a relaxed state in the EmotivBCI program, and (right, red) pull mental command training, where participants attempt to imagine a “pull” action while the Unity engine triggers object shrinkage through a Wizard-of-Oz approach. Each command was trained in 20 repetitions, with accepted trials saved to the user's training profile for later classification during the experiment}\label{fig:training-flowchart}
\end{figure}

The experimental software was developed in Unity (Unity Technologies, USA) using the Meta XR All-in-One SDK. Eye tracking was used to control a visual ray pointer and object hover state, rendered as a white line from the midpoint of the user's eyes to 500 meters in the forward direction. This ray cast triggered a scaling effect on interactable objects, causing them to grow to a fixed scale (0.2304m³) when hovered over and shrink back (0.18m³) when not. EEG signals were streamed into Unity through the Emotiv Cortex API. EEG calibration involved training two mental command classes: a neutral state (representing relaxed, unfocused brain activity) and a “pull” command associated with selection. During calibration, participants viewed objects that appeared and shrank in synchrony with their imagined action, providing feedback to reinforce consistent neural patterns (this was achieved through a Wizard-of-Oz approach in which the experiment administrator manually triggered the object to shrink seen in Figure \ref{fig:training-flowchart}). Once trained, the classifier output was integrated into the Unity selection loop: objects under gaze became eligible for interaction, and a detected pull command triggered selection. The EmotivBCI program handled training profiles, EEG noise sanitization, and classification of EEG artifacts.
To ensure synchronization across devices, event markers from Unity were transmitted to the EEG stream via the Cortex API, and system timestamps were aligned to the host computer's monotonic clock. Pilot testing verified timing precision within ±20 ms between modalities, sufficient for behavioral comparison across input conditions.

\subsection{Task}
\begin{figure}[ht]
\begin{center}
\includegraphics[width=15cm]{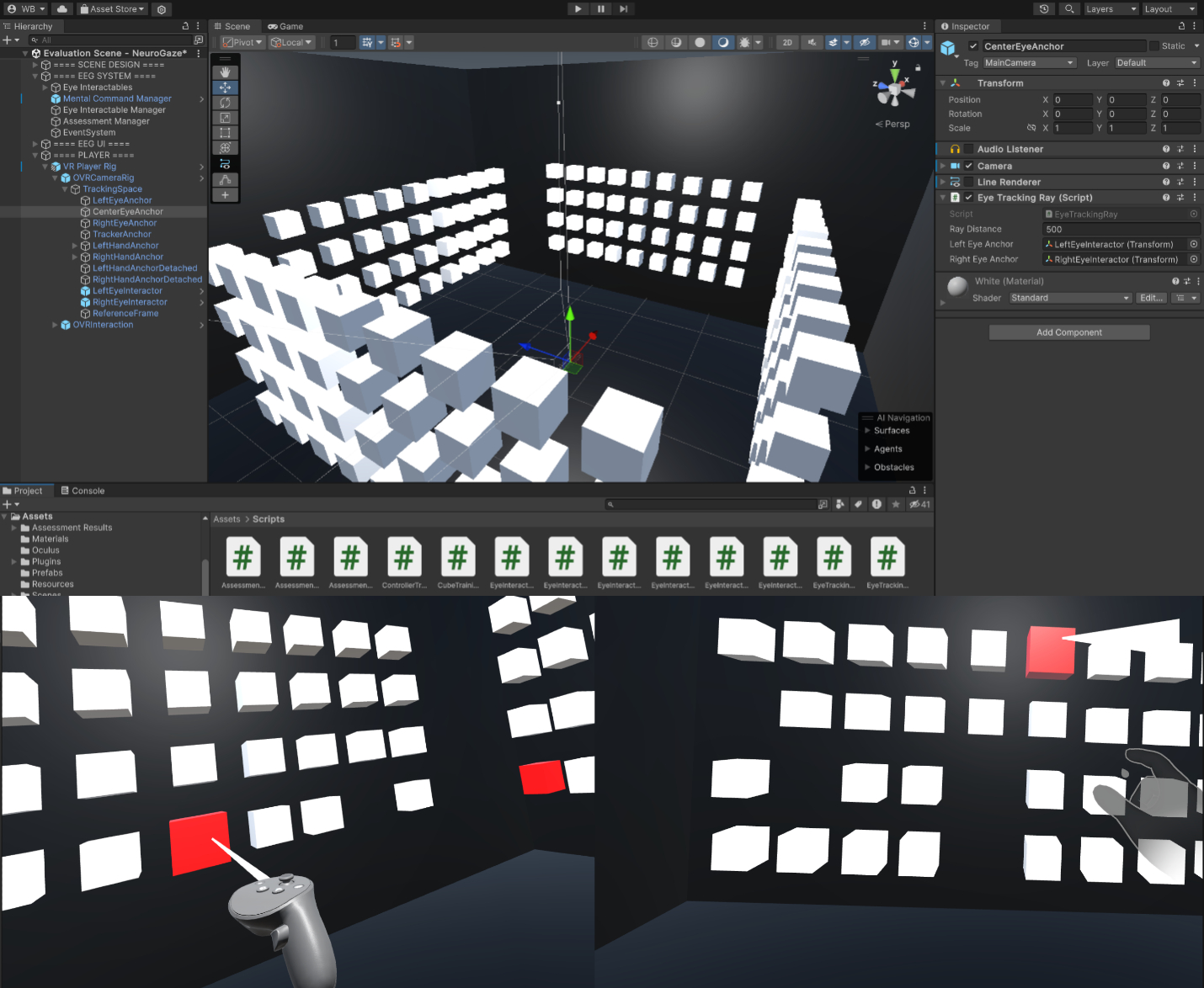}
\end{center}
\caption{Top: Unity editor view of the 360° object-selection task environment, with four cube arrays surrounding the participant. Bottom: Example participant perspectives during task execution. Left: VR Controller (VRC) condition; Right: Eye Gaze + Hand Gesture (EG+HG) condition.}
\label{fig:unity}
\end{figure}

Participants completed a 360° object-selection task in a virtual environment (VE). The environment consisted of four surrounding walls, each displaying a 4 x 9 array of white cubes (36 per wall; 144 total) as seen in Figure \ref{fig:unity}. 

At the start of each block, 12 cubes (three per wall) were designated as targets by turning red. Participants were instructed to select these targets as quickly and accurately as possible. When a target was successfully selected, it disappeared from the scene, and the block concluded once all targets had been cleared (Figure \ref{fig:unity}). The average distance from the user to each wall of cubes was approximately 2 m.

The task required participants to actively rotate their heads and bodies to engage with spatially distributed targets across the 360° field. Visual readiness feedback was provided by a scaling effect: when a user's gaze or pointing ray intersected a cube, it gradually increased in size (from ~0.56 m to ~0.62 m per side, corresponding to a 0.18 m³ to 0.23 m³ volume). Objects remained fixed in position and provided feedback only through this change in scale.

\subsection{Experimental Conditions}
Each participant completed the selection task under three input modalities:

VR Controllers (VRC). Participants used standard handheld VR controllers to interact with the virtual environment. A ray projected from the end of the controller was used to aim at targets, with selection confirmed via a trigger button press. This condition represented the conventional VR input method and served as the baseline for speed and precision.

Eye Tracking + Hand Gesture (EG+HG). Participants aimed by fixating on a target cube using the Quest Pro's integrated infrared eye-tracking system. Selection was confirmed with a pinch gesture detected by the headset's optical hand-tracking system. This approach provided gaze-driven aiming with explicit manual confirmation, similar to interaction paradigms employed in emerging augmented reality headsets.

NeuroGaze (NG). Participants aimed using eye gaze, with selection confirmed by an EEG-based “pull” mental command classified in real time by the Emotiv EPOC X headset. This condition enabled fully hands-free interaction through a hybrid brain-computer interface. The NeuroGaze system used a closed-loop control design, combining gaze-based ray casting with visual scaling feedback (grow/shrink cues) to indicate selection readiness and execution.
The order of conditions was randomized across participants to minimize order and learning effects.

\subsection{Measures}
Task performance was assessed using two primary behavioral measures. First, completion time was defined as the elapsed time (in milliseconds) between target onset and confirmed selection. This measure captured how long participants required to select all red interactable objects in the scene, with values exported directly from Unity environment logs. Second, error rate was defined as the proportion of incorrect or unintended selections relative to total trials, including both missed targets and incorrect object selections. Error data were compiled from block-level outcomes.

Subjective workload was evaluated after each condition using the NASA Task Load Index (NASA-TLX), which provides ratings across six subscales: Mental Demand, Physical Demand, Temporal Demand, Performance, Effort, and Frustration.  To derive the overall workload score, the subscales were combined according to Equation \ref{eq:nasatlx}.

\begin{equation}
\begin{split}
\text{NASA-TLX} &= \text{Mental Demand} + \text{Physical Demand} \\
&+ \text{Temporal Demand} + (7 - \text{Performance}) \\
&+ \text{Effort} + \text{Frustration}
\end{split}
\label{eq:nasatlx}
\end{equation}

Both aggregated NASA-TLX scores and individual subscale ratings were retained for analysis.

Finally, overall preference was captured through a post-experiment ranking task. After completing all three input conditions, participants ranked the modalities from most preferred (rank = 1) to least preferred (rank = 3). This ranking provided a simple comparative index of participants' subjective impressions of each input method.

\subsection{Analysis Plan}
Task completion time was analyzed with a repeated-measures design. Mauchly's test of sphericity was first applied; when violations were detected ($p < 0.001$), Greenhouse-Geisser corrections were used. A repeated-measures ANOVA was then conducted with Input Condition (VR Controllers, Eye+Pinch, NeuroGaze) as the within-subjects factor. Significant effects were followed up with Bonferroni-corrected pairwise t-tests. Effect sizes (partial $\eta^2$) were reported alongside significance values.

Error rates were analyzed similarly. Mauchly's test indicated that the assumption of sphericity was met ($p = 0.85$), so a repeated-measures ANOVA was conducted on average error counts. Post-hoc comparisons were performed with paired t-tests, and $\eta^2$ effect sizes were reported.

Subjective workload was evaluated using NASA-TLX ratings. Aggregated workload scores were compared across conditions using a Friedman test. Individual subscales (Mental Demand, Physical Demand, Temporal Demand, Performance, Effort, and Frustration) were analyzed with Wilcoxon signed-rank tests. Bonferroni correction was applied, setting the adjusted threshold for significance at $p < 0.003125$.

User preference rankings were analyzed with a Chi-squared test of independence to examine associations between input modality and rank position.
Across all analyses, 95\% confidence intervals were reported to provide interval estimates of effects. Visualizations were prepared to illustrate group-level distributions, including raincloud plots for completion time, bar plots for error rates, and radar charts for NASA-TLX subscales.

\section{Results}
The results are organized into three subsections corresponding to the main dependent measures: task completion time, error rate, and subjective workload. Statistical analyses were performed using repeated-measures designs with Condition (VR Controllers, Eye + Hand Gesture, NeuroGaze) as the within-subjects factor. All reported pairwise comparisons used Bonferroni-corrected p-values, and effect sizes are presented alongside significance values.

\subsection{Completion Time}
Task completion time differed significantly across input conditions. A repeated-measures ANOVA with Greenhouse-Geisser correction (due to sphericity violation, $\chi^2$(20) $= 29.22$, $p < 0.001$) showed a robust main effect of condition, $F(2, 19.77) = 97.62$, $p < 0.001$, $\eta_p^2$ $= 0.84$.

Participants completed the selection task fastest with VR Controllers (M = 9.25 s, SD = 4.15 s), followed by Eye + Hand Gesture (M = 15.02 s, SD = 5.30 s), and slowest with NeuroGaze (M = 29.23 s, SD = 2.25 s) (Figure \ref{fig:graphs}). Pairwise comparisons confirmed that both VR Controllers ($p < 0.001$) and Eye + Hand Gesture ($p < 0.001$) were significantly faster than NeuroGaze. VR Controllers were also significantly faster than Eye + Hand Gesture ($p < 0.001$).

This pattern indicates that while NeuroGaze enabled reliable hands-free selection, its current implementation introduced substantial latency compared to standard input methods.

\subsection{Error Rate}
Error rates differed significantly across input conditions. A repeated-measures ANOVA (Mauchly's test indicated sphericity was met: $\chi^2$(20) = 4.93, $p = 0.85$) revealed a main effect of condition, $F(2, 36) = 5.39$, $p = 0.009$, $\eta_p^2$ $= 0.23$.

On average, participants made the fewest errors with NeuroGaze (M = 2.25, SD = 1.08), followed by VR Controllers (M = 4.15, SD = 1.56) and Eye + Hand Gesture (M = 5.30, SD = 2.25) (Figure \ref{fig:graphs}). Pairwise contrasts showed significantly fewer errors in NeuroGaze compared with Eye + Hand Gesture ($p = 0.041$). Differences between NeuroGaze and VR Controllers ($p = 0.441$) and between VR Controllers and Eye + Hand Gesture ($p = 0.105$) were not significant.

\subsection{NASA-TLX}
\begin{figure}[ht]
\begin{center}
\includegraphics[width=17cm]{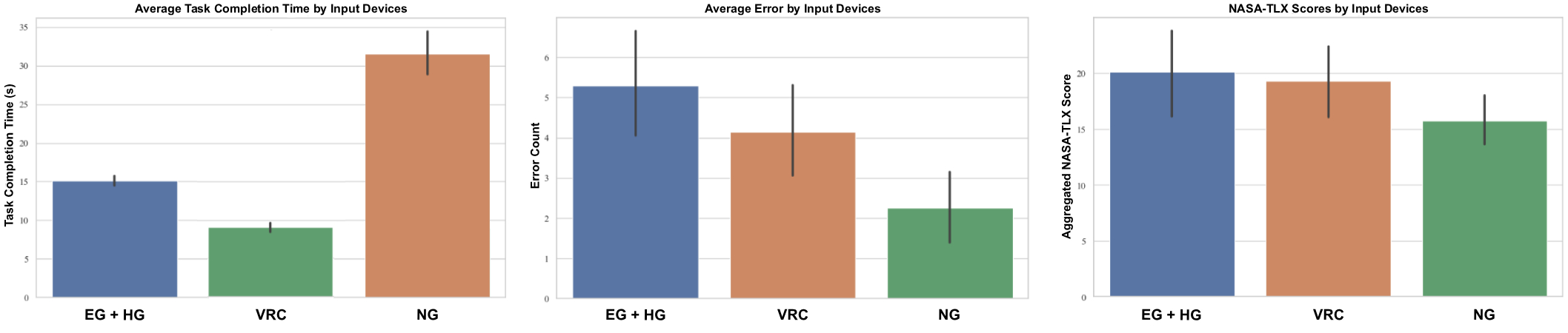}
\end{center}
\caption{Task performance and subjective workload across input devices. 
Left: Average task completion time (s). 
Middle: Average error count. 
Right: Aggregated NASA-TLX workload scores. 
Conditions are labeled as follows: EG+HG = Eye Gaze + Hand Gesture, VRC = VR Controllers, NG = NeuroGaze. 
Error bars represent 95\% confidence intervals.}
\label{fig:graphs}
\end{figure}

Subjective workload ratings from the NASA-TLX revealed differences across conditions, although patterns varied by subscale. Aggregated workload scores did not differ significantly between modalities (VR Controllers: M = 19.30; Eye + Hand Gesture: M = 20.10; NeuroGaze: M = 15.75), Friedman $\chi^{2}(2) = 0.29$, $p > 0.05$ (Figure \ref{fig:graphs}).

When subscales were examined individually using Wilcoxon signed-rank tests with Bonferroni correction ($p < 0.003125$), more specific distinctions emerged. Physical Demand was lowest for NeuroGaze (M = 1.3), significantly lower than VR Controllers (M = 3.3, $p = 0.002$). The comparison with Eye + Hand Gesture (M = 3.6) trended in the same direction ($p = 0.006$) but did not survive correction. No difference was observed between VR Controllers and Eye + Hand Gesture ($p = 0.89$).

Temporal Demand also differed: NeuroGaze (M = 3.2) was rated significantly less demanding than both VR Controllers ($p = 0.001$) and Eye + Hand Gesture ($p = 0.002$).
For Mental Demand, no significant differences were found (NeuroGaze M = 3.2; VR Controllers M = 3.3; Eye + Hand Gesture M = 2.6). Similarly, Performance, Effort, and Frustration ratings did not differ significantly after correction.

\subsection{User Preference}
After completing all three conditions, participants ranked the input modalities by overall preference. NeuroGaze was most often ranked first (10 participants), followed by VR Controllers (5) and Eye + Hand Gesture (5). Intermediate rankings were more evenly distributed (VR Controllers: 8; NeuroGaze: 6; Eye + Hand Gesture: 6). Least-preferred rankings were most frequently assigned to Eye + Hand Gesture (9 participants), followed by VR Controllers (7) and NeuroGaze (4). 

A chi-squared test of independence revealed no significant association between input modality and preference ranking, $\chi^2$($2, N = 20$) $= 4.8$, $p = 0.31$.

Qualitative feedback provided additional context. VR Controllers were praised for their speed and familiarity. Eye + Hand Gesture was described as intuitive but often unreliable, with several participants noting difficulty executing pinch gestures. NeuroGaze was appreciated for its novelty and hands-free interaction, though some participants reported discomfort from wearing both headsets and noted slower response times.

\section{Discussion}
\subsection{Interpretation}
The present findings highlight both the strengths and limitations of current VR input modalities. Handheld controllers remain the benchmark for speed, with participants consistently achieving the fastest completion times. This reflects both the maturity of the technology and its optimization for rapid, precise selection tasks. Eye tracking with pinch gestures (EG+HG) occupied an intermediate position, offering more intuitive aiming than controllers but at the expense of occasional mis-selections from incomplete or unrecognized gestures.

In contrast, NeuroGaze yielded fewer errors overall, but this advantage came at the cost of substantially slower task completion times. Accuracy appears to result less from superior input fidelity than from participants’ conservative pacing, further amplified by the processing delay inherent in EEG classification, a clear example of the classic speed-accuracy tradeoff. Thus, while NG does not yet rival existing VR inputs on raw performance metrics, it provides a viable, hands-free alternative that emphasizes ergonomic accessibility \citep{101371,6451193} rather than speed. Rather than competing with controllers in high-speed applications, NG is better suited to daily-activity contexts where comfort, inclusivity, and error minimization are paramount.

Workload ratings reinforce this role. NeuroGaze reduced perceived physical and temporal demand compared with conventional inputs, while overall cognitive workload remained comparable across modalities. This suggests that, despite slower performance, NG may lower effort and fatigue, making it especially relevant for accessibility and long-duration use cases.

\subsection{Contribution}
This study provides the first systematic evaluation of a consumer-grade hybrid EEG and eye-tracking system benchmarked against established VR input modalities in a fully immersive environment. Prior work on gaze-EEG interaction has largely relied on laboratory-grade hardware or 2D desktop displays, limiting ecological validity and applicability to everyday contexts. By deploying NeuroGaze with widely available devices—the Meta Quest Pro and Emotiv EPOC X—this study shows that hybrid brain-computer interfaces are no longer confined to specialized laboratories and can be assessed under conditions closer to daily VR use.

Benchmarking NeuroGaze against two established input modalities (controllers and gaze + pinch) further clarified its comparative strengths and weaknesses. While slower than conventional inputs, NeuroGaze offers a tangible ergonomic benefit, demonstrated by lower physical demand ratings and fully hands-free operation. These qualities suggest that the system is not a competitor to controllers in time-sensitive or performance-critical contexts, but rather a complementary modality where accessibility, comfort, and reduced fatigue are prioritized.
The most promising applications of NeuroGaze may therefore lie in daily-activity  and accessibility-oriented scenarios that demand sustained interaction without physical strain \citep{Sellers01102010}. Examples include VR-based rehabilitation, training for individuals with motor impairments, or prolonged use cases where repetitive arm or hand motions become burdensome. By reframing the role of BCIs away from speed competition and toward ergonomic inclusivity, this study contributes to a broader vision of BCIs as practical tools for everyday human-computer interaction.

\subsection{Limitations}
Several limitations of the present study should be acknowledged. First, the sample size was modest (N = 20), which restricts the generalizability of the findings and reduces the statistical power to detect more subtle effects. While sufficient for an initial proof-of-concept, larger studies will be needed to establish more robust estimates of performance and variability across different populations.
Second, the task design employed static targets arranged across four walls. This setup provided consistency across conditions but does not capture the more dynamic and unpredictable environments in which VR interactions typically occur. Future work should examine performance in tasks involving moving or context-sensitive stimuli to evaluate real-world applicability.
Third, the EEG calibration procedure incorporated a Wizard-of-Oz component in which feedback was artificially reinforced to improve classifier training. Although the actual task relied on trained classifiers, this approach may have inflated participants' perception of system reliability during calibration.
A further limitation arises from the use of consumer-grade EEG hardware (Emotiv EPOC X), which is constrained by relatively low signal-to-noise ratios. In practice, this restricted the system to a binary command scheme (neutral vs. pull), as attempts to distinguish more complex mental commands (e.g., push and pull) would have introduced substantial classification errors. Relatedly, reliance on consumer-grade EEG made the system more susceptible to artifacts such as blinking and head movement, and the limited spatial resolution reduced the sophistication of neural information that could be leveraged.

Finally, ergonomic incompatibility between the EEG headset and the Meta Quest Pro contributed to discomfort during extended use. While employing commercially available devices strengthens ecological validity, these hardware limitations necessarily constrained both the fidelity of neural input and the overall user experience.

\subsection{Future Work}
Several avenues for future development emerge from the present findings. A key priority is the reduction of system latency. NeuroGaze's slower performance relative to traditional input methods reflects both the computational overhead of EEG signal classification and the conservative thresholds used to minimize false activations. Advances in machine learning and signal processing—such as adaptive filtering, transfer learning across users, and real-time artifact rejection—may help reduce response times while maintaining accuracy, thereby improving the practical viability of hybrid BCI input.
Another promising direction involves adaptive multimodal switching, in which NeuroGaze could dynamically integrate with conventional controllers or gesture-based systems. For example, users might rely on EEG+gaze input for sustained, low-effort interaction but seamlessly transition to controller-based input when speed or fine-grained control is required. Such hybrid workflows would leverage the strengths of each modality and broaden the contexts in which BCIs are practical.
Beyond EEG alone, integration with complementary biosignals represents a further step forward. Modalities such as functional near-infrared spectroscopy (fNIRS), electromyography (EMG), or pupillometry could provide additional channels for intent detection and cognitive-state monitoring. Combining signals could improve classification robustness, reduce reliance on single noisy channels, and support more complex command vocabularies than binary EEG triggers allow.
Finally, future studies should move beyond healthy young adults to evaluate NeuroGaze in accessibility scenarios. Populations with motor impairments, fatigue-related conditions, or limited hand mobility stand to benefit most from hands-free BCI interaction. Assessing usability, comfort, and performance in these groups will be essential for determining NeuroGaze's translational potential in rehabilitation, assistive technology, and daily activity contexts.

\section{Conclusion}
This study introduced and evaluated NeuroGaze, a hybrid EEG and eye-tracking interface implemented with consumer-grade hardware in an immersive VR environment. Compared to conventional VR controllers and gaze+pinch interaction, NeuroGaze enabled reliable, fully hands-free object selection, though at the cost of slower task completion times. The results reflect a classic speed-accuracy tradeoff: participants made fewer errors with NeuroGaze, but this advantage stemmed largely from more deliberate pacing rather than inherently superior input fidelity.
Despite these performance constraints, NeuroGaze demonstrates clear ergonomic and accessibility promise. By reducing physical demand and eliminating the need for handheld controllers, it extends VR interaction beyond speed-driven contexts toward scenarios where comfort, inclusivity, and reduced fatigue are prioritized. Rather than serving as a replacement for controllers in time-critical tasks, NeuroGaze should be considered a complementary modality for daily activities, rehabilitation contexts, and fatigue-sensitive environments where minimizing physical effort is critical.
Taken together, these findings establish the feasibility of hybrid EEG+gaze interaction in immersive VR using readily available consumer devices. More broadly, they highlight the potential of consumer-grade BCIs not as direct competitors to established input methods, but as enablers of more inclusive and adaptable human-computer interaction.

\section*{Acknowledgments}
Portions of this research were previously included in the Master's thesis of Wanyea Barbel, archived in the University of Central Florida STARS Digital Repository. The present manuscript is a reformatted and condensed version of that work \citep{Barbel2024}.

\section*{Conflict of Interest Statement}
The authors declare that the research was conducted in the absence of any commercial or financial relationships that could be construed as a potential conflict of interest.

\section*{Data Availability Statement}
The datasets analyzed for this study can be found in the GitHub repository here: https://github.com/Wanyea/NeuroGaze.

\bibliography{refs}
\end{document}